# Electron resonant tunneling through InAs/GaAs quantum dots embedded in a Schottky diode with an AlAs insertion layer

Jie Sun[a)], **Peng Jin, Chang Zhao, Like Yu, Xiaoling Ye, Bo Xu, Yonghai Chen, and Zhanguo Wang[b)]**

*Key Laboratory of Semiconductor Materials Science, Institute of Semiconductors, Chinese Academy of Sciences, P. O. Box 912, Beijing 100083,* ***China***

Molecular beam epitaxy is employed to manufacture self-assembled InAs/GaAs quantum dot Schottky resonant tunneling diodes. By virtue of a thin AlAs insertion barrier, the thermal current is effectively reduced and electron resonant tunneling through quantum dots under both forward and reverse biased conditions is observed at relatively high temperature of 77K. The ground states of quantum dots are found to be at ~0.19eV below the conduction band of GaAs matrix. The theoretical computations are in conformity with experimental data.

**a) Electronic mail: albertjefferson@sohu.com**
**Present at: Solid State Physics, Lund University, P. O. Box 118, SE-221 00 Lund, Sweden**
**b) Electronic mail: zgwang@red.semi.ac.cn**

## I. INTRODUCTION

In the past two decades, self-organized InAs/GaAs quantum dots (QDs) grown by Stranski-Krastanow (S-K) method have been intensively investigated[1], not only due to their unique properties of "artificial atoms", but due to their possible device applications as well. Although InAs/GaAs QDs are often inserted into Schottky diodes to study their electronic structure[2,3], electron resonant tunneling through the QDs in these Schottky diodes is very rarely focused on[4-6]. Usually, quantum-dot resonant tunneling signals in current-voltage (I-V) characteristics can only be obtained at extremely low temperature (typically ≤4.2K), either in Schottky resonant tunneling diodes[4] (RTDs) or in double barrier RTDs with two-side Ohmic contacts[7,8]. The main reason is, however, the signals are obscured by thermal current. By virtue of electron beam lithography technique, resonant tunneling via an individual QD in Schottky RTD can be observed at ~130K[5]. When very small mesas are fabricated (0.5μm square in Ref. 5), their peripheral regions are not active but depleted and that can to some extent increase the ratio of resonant tunneling current to hot current. Li and Wang's[6] experiments are executed at 77K, but they have not clarified why the detected resonant tunneling peaks do not originate from bound states of $Al_{0.3}Ga_{0.7}As/GaAs$ superlattice structures in their devices. To the best of our knowledge, one can observe electron resonant tunneling through QD Schottky RTDs only in one voltage direction[4-6]. In the present work, we will exhibit resonant tunneling semaphores under both forward and reverse biased conditions at relatively high temperature of 77K through reducing the thermal current in InAs/GaAs quantum-dot Schottky RTDs by inserting a thin AlAs barrier layer. Our theoretical calculations coincide with experimental results very well.

## II. EXPERIMENT

The QD Schottky RTDs, schematically shown in Fig. 1, are grown by solid source molecular beam epitaxy (MBE, RIBER 32P) on $n^+$-doped GaAs (100) substrate with Si of $2\times10^{18}cm^{-3}$. On the $600^oC$-grown-$n^+$-GaAs buffer layer ($n^+=2\times10^{18}cm^{-3}$), we grow 2nm undoped GaAs spacer layer and 4nm undoped AlAs insertion layer. Then, the wafer temperature is lowered to $500^oC$ and 3 monolayers (ML) undoped GaAs prelayer is deposited. That aims to prevent the direct contact of aluminum-related defects with InAs QDs. At the same temperature, nominally 1.5 ML InAs are deposited at a very slow velocity of 0.0038ML/s (the In delivery is cycled in 1s of evaporation followed by 3s of interruption until the given InAs thickness is reached) and 2 min post-growth annealing[9] is processed. The growth of our RTD structures is terminated by deposition of twofold 5nm undoped GaAs cap layers at $500^oC$ and $600^oC$ respectively. After the sample is removed from the MBE chamber, alloyed Au/Ge/Ni back Ohmic contact is prepared. 150nm silicon dioxide is deposited onto its facade and 5μm×5μm windows are photolithographed therein to confine the conduction current. Ti/Au Schottky contact is fabricated via these windows by depositing 220nm Ti and 150nm Au through vacuum evaporation at room temperature. Finally, the wafer is cut into pieces of 0.5mm×0.5mm to manufacture RTDs. The I-V properties are measured by KEITHLEY 4200-SCS/F semiconductor characterization system at 77K. The wafer photoluminescence (PL) is carried out at 77K using the 514.5nm line of an $Ar^+$ laser for excitation.

## III. THEORY

The conduction band profile of our Schottky RTDs can be calculated analytically. Since there are intrinsic layers between metal and n-type semiconductors, the case is more complicated than standard Schottky junction illustrated in semiconductor physics textbooks. Fig. 2 shows the energy band diagram of the RTD in thermal equilibrium, where $x_d$ is the depletion layer width, and $-V_0$, $-\Phi$, $\Phi_n$ are electric potentials at x=-a, x=0 and $x>x_d$, respectively ($V_0$, $\Phi$, $\Phi_n>0$). **The origin of the coordinates, x=0, is selected at the interface between 2nm undoped GaAs and $n^+$ GaAs buffer layer.** $E_0$ is the differential value between the binding energy state in InAs QD and the conduction band edge of GaAs. The AlAs layer is not exhibited in Fig. 2 for simplification reason. The electric-field and potential distribution of the RTD can be obtained by solving the Poisson's equation. Under the **so called** abrupt-junction approximation (pp. 163 in Ref. 10) that the space charge density $\rho_s \approx qN_D$ for $0 \le x \le x_d$, and $\rho_s=0$ and $dV/dx \approx 0$ for $x>x_d$, where q is elementary charge, V is electric potential and $N_D$ is donor concentration, the Poisson's equation can be written as follows:

$$\frac{d^2V}{dx^2} = \begin{cases} -\frac{qN_D}{\varepsilon_r \varepsilon_0} & (0 \le x \le x_d) \\ 0 & (x > x_d) \\ 0 & (x < 0). \end{cases} \quad (1)$$

Here, $\varepsilon_r$ and $\varepsilon_0$ are permittivity in GaAs and in vacuum respectively. Our boundary condition is the continuity of electric displacement, that is, $\varepsilon_1 E_1=\varepsilon_2 E_2$, where E represents electric field (pp. 146 in Ref. 10). In our case, $\varepsilon_1=\varepsilon_2=\varepsilon_r$, and thus $E_1=E_2$. The boundary conditions are given by

$$E(0)=\frac{\Phi-V_0}{a};\quad E(x_d)=0. \tag{2}$$

Therefore,

$$E(x)=-\frac{dV(x)}{dx}=\begin{cases}\frac{qN_D}{\varepsilon_r\varepsilon_0}x+\frac{\Phi-V_0}{a}=\frac{qN_D}{\varepsilon_r\varepsilon_0}x-\frac{qN_D}{\varepsilon_r\varepsilon_0}x_d & (0\le x\le x_d)\\ 0 & (x>x_d)\\ \frac{\Phi-V_0}{a}=-\frac{qN_D}{\varepsilon_r\varepsilon_0}x_d & (x<0).\end{cases} \tag{3}$$

The last equation in that bracket also gives the relationship between Φ and $x_d$:

$$\Phi=V_0-\frac{aq\,N_D}{\varepsilon_r\varepsilon_0}x_{d.} \tag{4}$$

Considering the boundary condition

$$V(0)=-\Phi, \tag{5}$$

we can get the electric potential distribution from Eq. (3):

$$V(x)=\begin{cases}-\frac{qN_D}{2\varepsilon_r\varepsilon_0}x^2-\frac{\Phi-V_0}{a}x-\Phi=\frac{qN_D}{\varepsilon_r\varepsilon_0}(-\frac{1}{2}x^2+x_dx+ax_d)-V_0 & (0\le x\le x_d)\\ \Phi_n & (x>x_d)\\ \frac{V_0-\Phi}{a}x-\Phi=\frac{qN_D}{\varepsilon_r\varepsilon_0}x_dx+\frac{aqN_D}{\varepsilon_r\varepsilon_0}x_d-V_0 & (x<0).\end{cases} \tag{6}$$

When bias V is applied to the RTD, we have $V(x_d)=\Phi_n-V$. That is

$$\frac{qN_D}{\varepsilon_r\varepsilon_0}(\frac{1}{2}x_d^2+ax_d)-(\Phi_n-V+V_0)=0 \tag{7}$$

and we get the relation between $x_d$ and the bias V:

$$x_d=\sqrt{a^2+\frac{2\varepsilon_r\varepsilon_0}{qN_D}(\Phi_n+V_0-V)}-a. \tag{8}$$

Now we will focus on at what bias electrons will resonant tunnel through InAs QDs. At forward bias, **when resonance takes place**, the Fermi level $E_f$ of $n^+$ GaAs should be brought into alignment with InAs QD energy state and the following conditions must be satisfied:

$$-(\frac{V_0-\Phi}{a}x_{QD}-\Phi)-E_0=V_{For}, \tag{9}$$

where $x_{QD}$ is the coordinate of InAs QDs. Simultaneously with Eqs. (4) and (8), we know that the bias V is determined by the root of this equation:

$$\begin{aligned}&\varepsilon_r\varepsilon_0 V_{For}{}^2+2\left[qN_D x_{QD}(x_{QD}+a)-\varepsilon_r\varepsilon_0(V_0-E_0)\right]V_{For}+\\&\varepsilon_r\varepsilon_0(V_0-E_0)^2-2qN_D(x_{QD}+a)\left[a(\Phi_n+E_0)+x_{QD}(\Phi_n+V_0)\right]=0.\end{aligned} \tag{10}$$

Similarly, at reverse bias, there are

$$-(\frac{V_0-\Phi}{a}x_{QD}-\Phi)-E_0=0, \tag{11}$$

and

$$V_{\mathrm{Re}v}=V_0+\Phi_n-\frac{\varepsilon_r\varepsilon_0(V_0-E_0)^2}{2qN_D(a+x_{QD})^2}-\frac{a(V_0-E_0)}{a+x_{QD}}. \tag{12}$$

Summarily, we have attained the electric-field and potential distribution in as-manufactured Schottky RTDs by means of solving Eq. (1). We have also got the conversion relations between $\Phi$, $x_d$ and between $x_d$, V (bias). $V_{For}$ and $V_{Rev}$ can be computed for a given $E_0$ value. Those developed formulae will play key roles thereinafter.

## IV. RESULTS AND DISCUSSION

When we prepare the active region of Schottky RTDs, although the InAs thickness is below critical value in S-K growth mode, formation of self-assembled QDs can still be achieved[9]. Low deposition rate, growth interruption and post-growth annealing all help to equilibrate the wafer surface by enhancing the migration of In adatoms[11] and consequently produce high quality QDs. The sheet density of QDs is $\sim 1.2\times10^{10}cm^{-2}$ as determined from plan-view transmission electron microscope (TEM) measurements [Fig. 3, **g**=(220)]. The QDs here are not as sparse as the situation described in Ref. 9. That is because on AlAs matrix material one usually gets dense dots[12]. The QD width is estimated to be 20-40nm. Fig. 4 is the 77K PL spectrum of our material. Three peaks are detected at ~0.952eV, ~1.112eV and ~1.313eV, which are attributed to photoluminescence from the ground states, first excited states and second excited states of the InAs/GaAs quantum dots respectively. Another peak situated at ~1.520eV originates from GaAs bulk material.

As is similar to previous studies[5,6], the resonant tunneling features can not be obtained from all devices. The mechanism of this, however, is still not understood. In some devices, the experimental I-V curve shows electron resonant signals at positive biased condition whereas it displays no structures at negative bias [Fig. 5(a)]. The small peak located at ~0.42V is caused by electron resonant tunneling via the ground states of QDs. Rarely, resonant tunneling in both scan directions can be observed (~0.48V and ~-0.81V), as is shown in Fig. 5(b). The spreading of I-V peaks is because of the size undulation and lateral coupling of QDs[13]. All current-voltage measurements are performed at liquid nitrogen temperature.

Now we will furnish explanations for the experimental findings using the theory developed in section III of this article. According to our RTDs, parameters in the calculation are defined as follows: $a=1.73\times10^{-8}$m, $x_{QD}=-7.08\times10^{-9}$m, $\varepsilon_r=12.4$(GaAs), $\varepsilon_0=8.854\times10^{-12}$F/m, $q=1.602\times10^{-19}$C, $N_D=2\times10^{24}$m$^{-3}$. Schottky barrier height $qV_0$ of Ti-GaAs contact is 0.83eV[14], and $\Phi_n$ is estimated to be 0.078V at 77K[15] (approximately 3kT, where k is Boltzmann constant). The Γ-valley conduction band offset of GaAs/AlAs heterostructure is around 1.03eV (pp. 5855 in Ref. 16). Thus, the position of ground eigenstate of QDs, $E_0$, is determined to be ~0.19eV by iterative computation of Eq. (10) and (12). In that case, $V_{For}\approx0.4$V and $V_{Rev}\approx-0.8$V. Other theoretically calculated parameters while resonant tunneling at forward and reverse voltages are summarized in Fig. 6(a) and (b). Obviously, Fig. 6 gives **the** good fit to the data in Fig. 5.

In **the Schottky** junction, the hot current is mainly dominated by the thermionic emission procedure[17]. So, in the mass, the higher the barrier is, the lower the current will be. We consequently calculated the gap between AlAs barrier top and the Fermi level of the emitter in our device at diverse biases. The difference between Ti-GaAs Schottky barrier top and the emitter Fermi energy in a common structure without AlAs insertion layer is also reckoned and summarized in Fig. 7. By comparison, we learn that the AlAs layer serves as an effective barrier for hot electrons, particularly when the device is positively biased. It is for this reason that we can observe resonant current in both directions at a relatively high temperature of 77K. In Fig. 7, we have also computed the extension of depletion region **in our devices** at various voltages. Even at high forward bias such as 0.8V, the width of depletion region is as large as 2.01nm. In respect that the total width of intrinsic layers in our RTDs is only 17.3nm, **the depletion region here is thus nonnegligible**, let alone the situation when the devices are negatively biased. Thereby, we argue that ignoring the depletion region in the schematic diagram of device energy band in aforetime researches[4,5] seems to be inappropriate. In fact, the depletion region is a high-resistance district. When the Schottky RTD is reversely biased, the resonant tunneling current has to wade across the much elongated depletion region (~21nm in our case) to arrive at the collector. That can explain why the resonant current at negative voltage is either extinguished[4,5] or not easy to be observed (this present work). Although in Ref. 6 the authors have reported reverse resonant tunneling current, **they** can not exclude the possibility that the signals might come from superlattice structures in their devices and can not interpret why the sign of forward resonant current is unseen.

## V. CONCLUSIONS

To sum up, we have successfully fabricated Schottky quantum-dot RTDs using MBE and simple device craft. By adding a thin AlAs barrier to the device, we have detected electron resonant tunneling current via the ground states of self-assembled

InAs/GaAs QDs under both forward and reverse biased conditions at 77K. The position of that energy level is at ~0.19eV below the conduction band of GaAs material. Reverse resonant current is difficult to be found because of the high resistance of the enlarged depleted layer. Our theory fits all the experimental data very well. It is expected that such achievements may contribute to a more thoroughgoing understanding of resonant tunneling through QDs.

## ACKNOWLEDGMENTS

We would like to thank **Professors Zhongli Liu and Jinlei Wu** for their assistance. Part of this work is finished with the help from Mrs. Chunli Yan. The financial aid from Special Funds for Major State Basic Research Project of China **(Nos 2000068303 and 2002CB311905)** is acknowledged.

## *Figure captions*

Fig. 1. Schematic illustration of the sample structure. The direction of electron flow at positive voltage is indicated by arrow.

Fig. 2. Conduction band profile of the Schottky RTDs at zero bias. The original point of x-axis is at the interface between the intrinsic layers and $n^+$-GaAs buffer layer. The Fermi level of metal is selected as the null point of electric energy. The AlAs layer is not shown in order to make the picture more simplex.

Fig. 3. A typical plan-view TEM image of the active region in as-grown device.

Fig. 4. 77K PL spectrum of the sample measured before it is made into RTDs.

Fig. 5(a). 77K I-V curve of a Schottky RTD which shows resonant signal at forward bias.

Fig. 5(b). 77K I-V curve of a Schottky RTD which shows resonant signal at both forward and reverse biases.

Fig. 6(a). Calculated energy band profile of the RTDs when resonant tunneling at 0.4V. Electric potentials of several black round spots are also given.

Fig. 6(b). Calculated energy band profile of the RTDs when resonant tunneling at -0.8V. Electric potentials of several black round spots are also given.

Fig. 7. Left: computed energy difference between barrier top and Fermi level at various voltages of Schottky RTDs with (square) or without (circle) AlAs insertion layer. Right: width of the depletion region in the Schottky RTDs with AlAs insertion layer at different biases (triangle symbol).

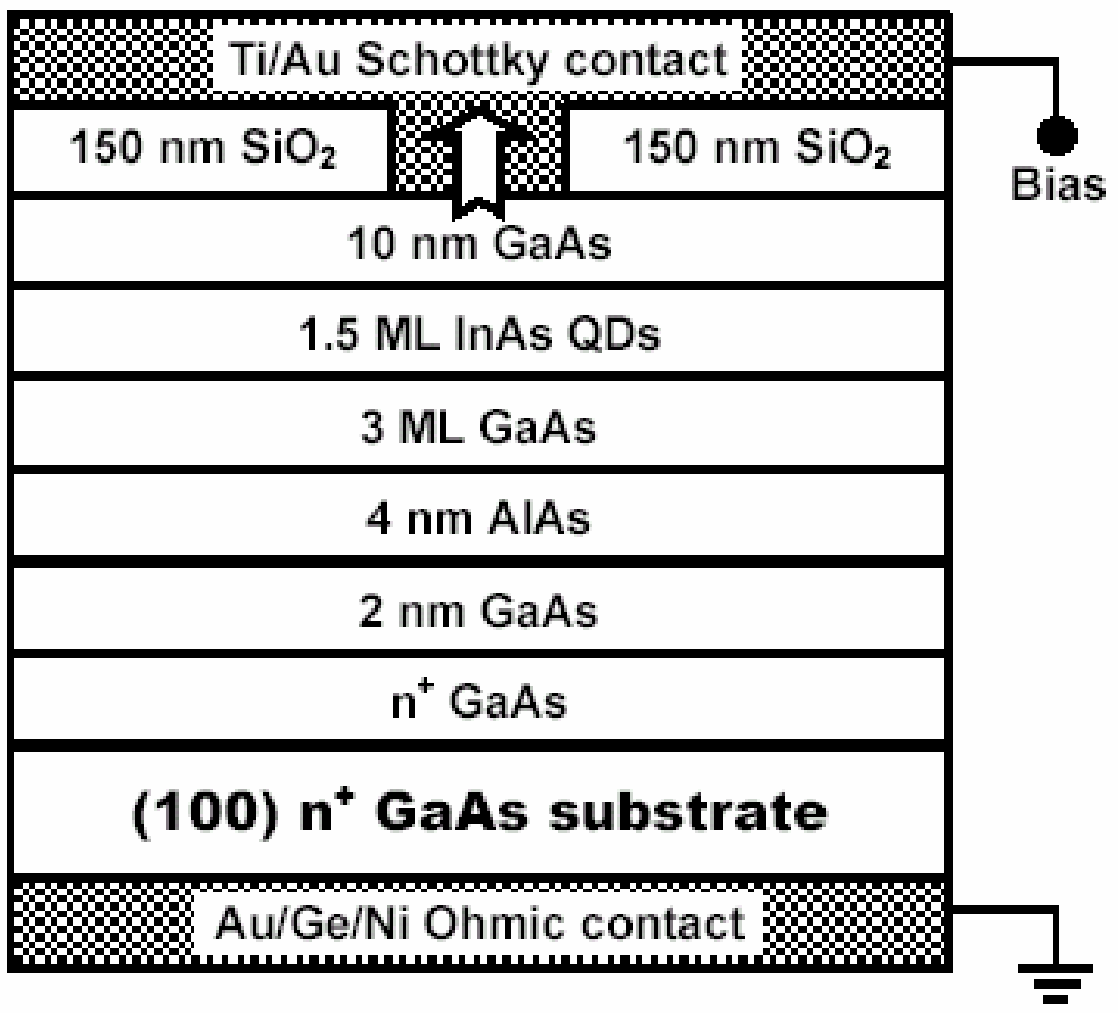

**Figure 1**

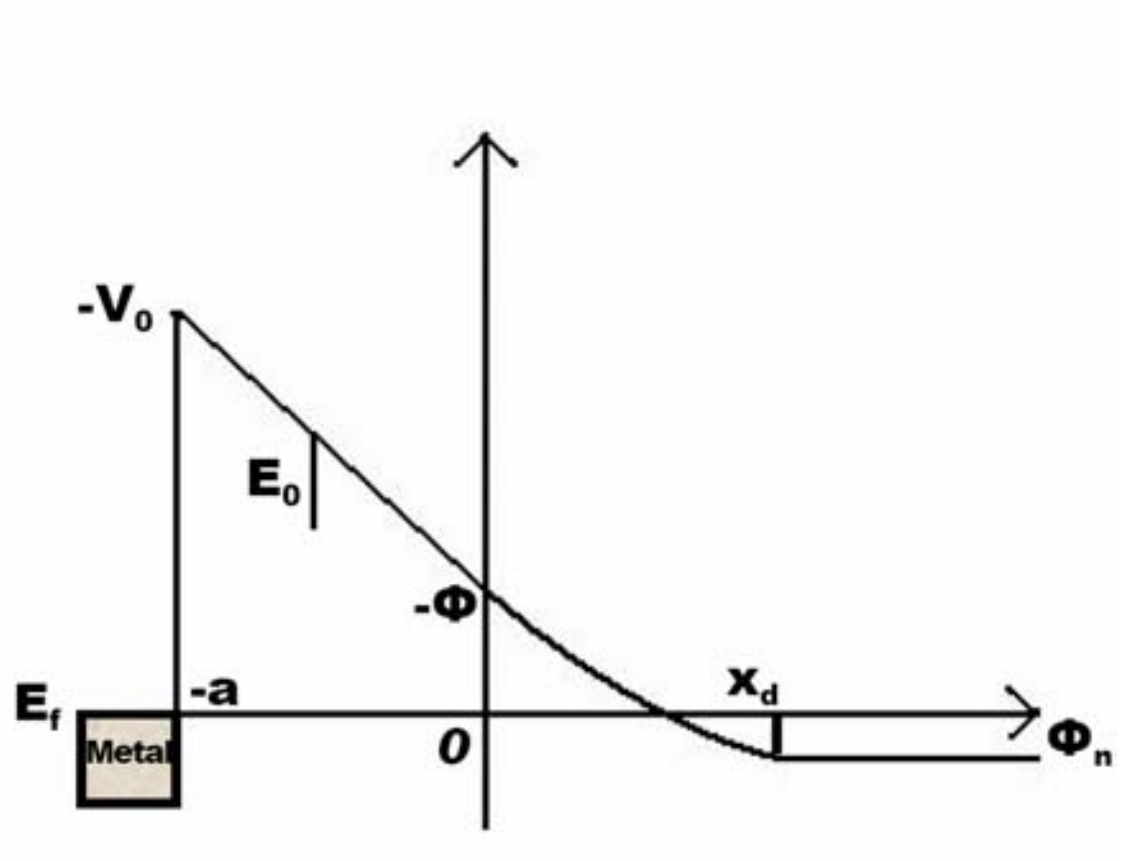

**Figure 2**

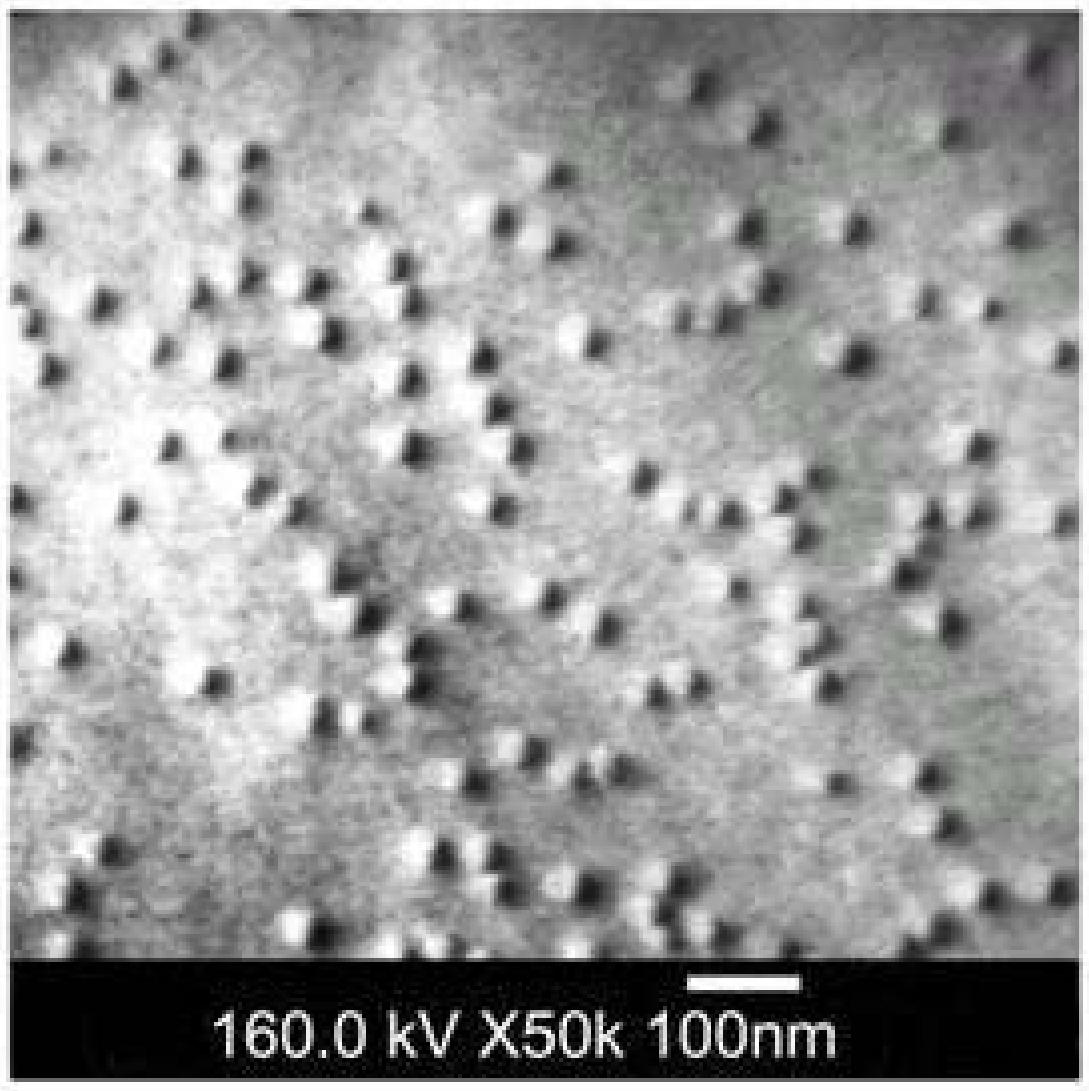

**Figure 3**

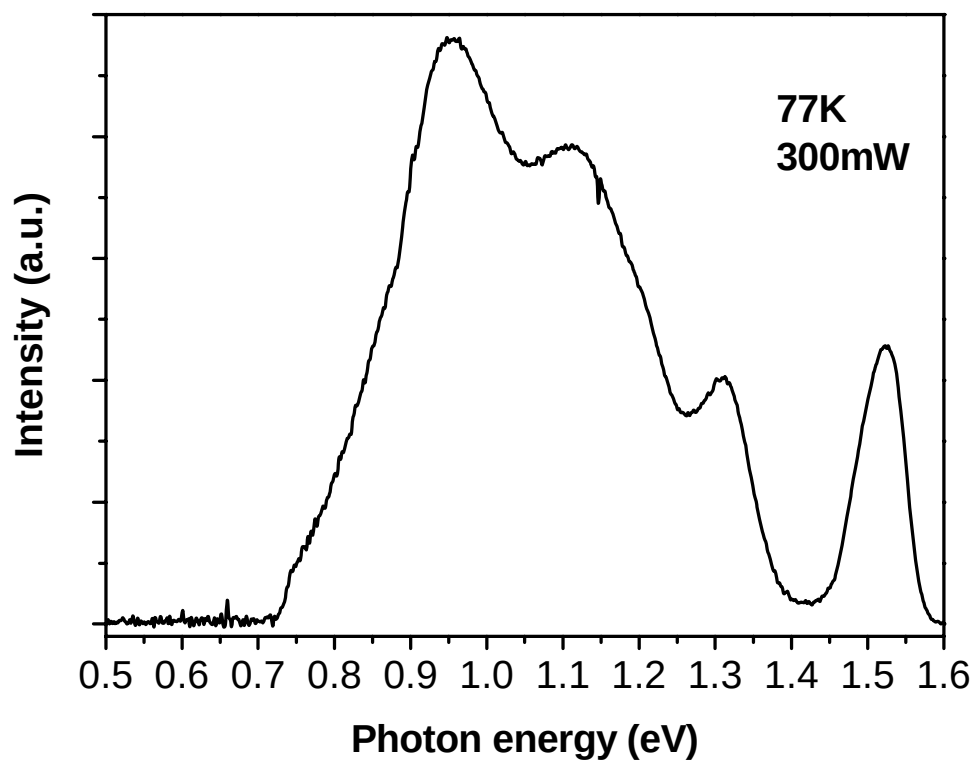
77K
300mW
Intensity (a.u.)
0.5 0.6 0.7 0.8 0.9 1.0 1.1 1.2 1.3 1.4 1.5 1.6
Photon energy (eV)

**Figure 4**

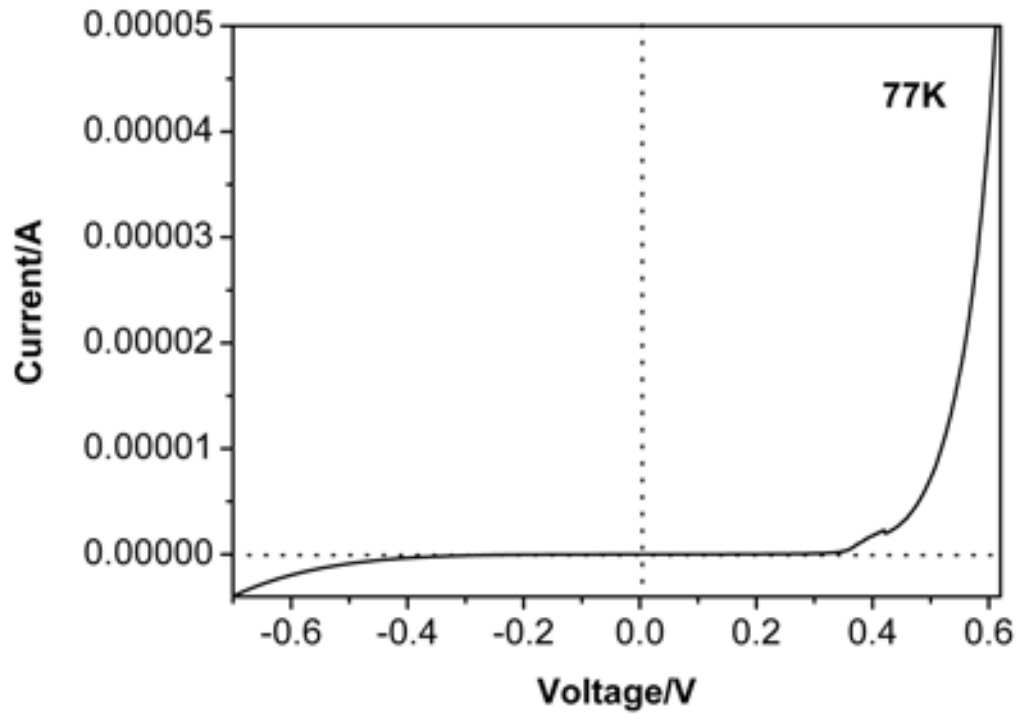

**Figure 5(a)**

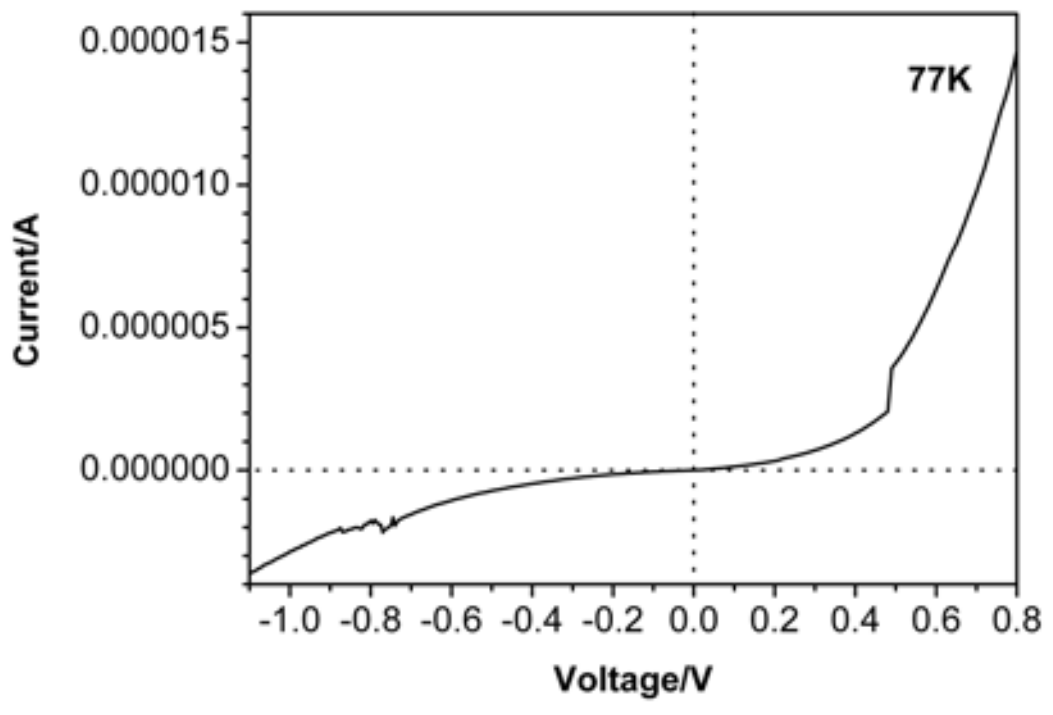

**Figure 5(b)**

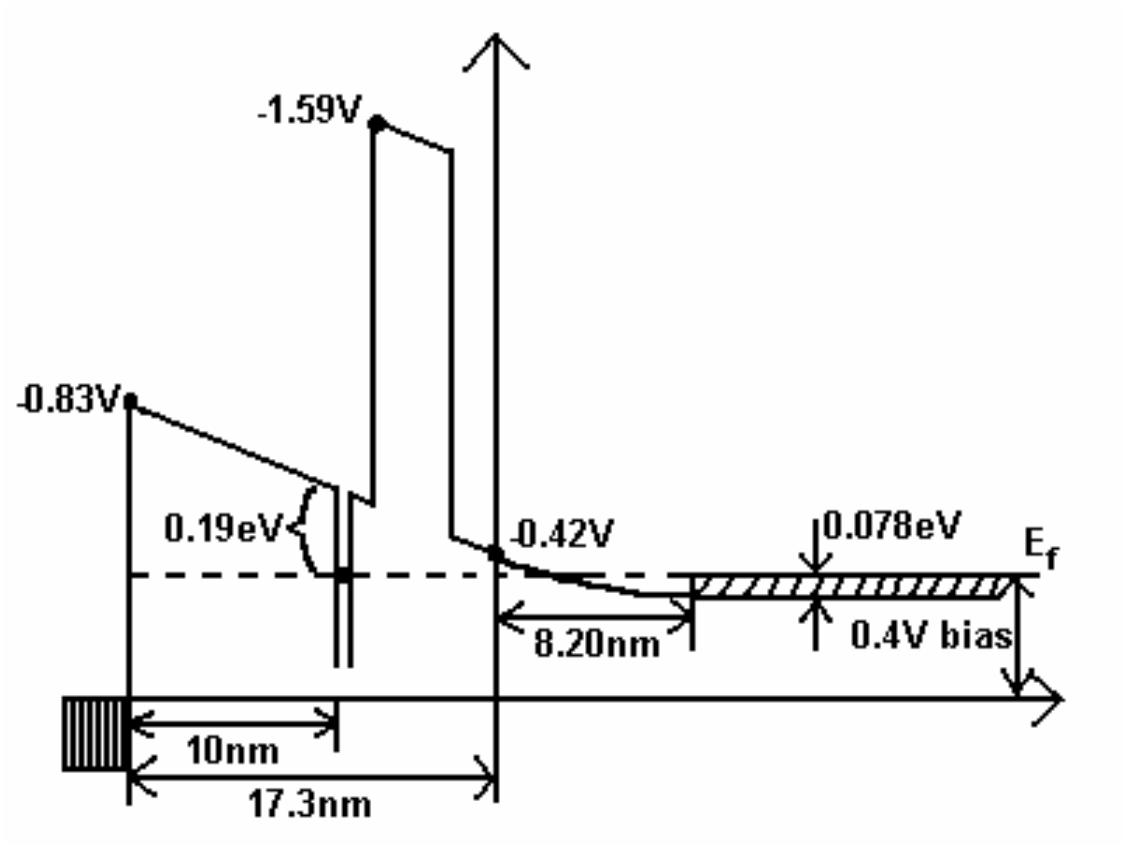

**Figure 6(a)**

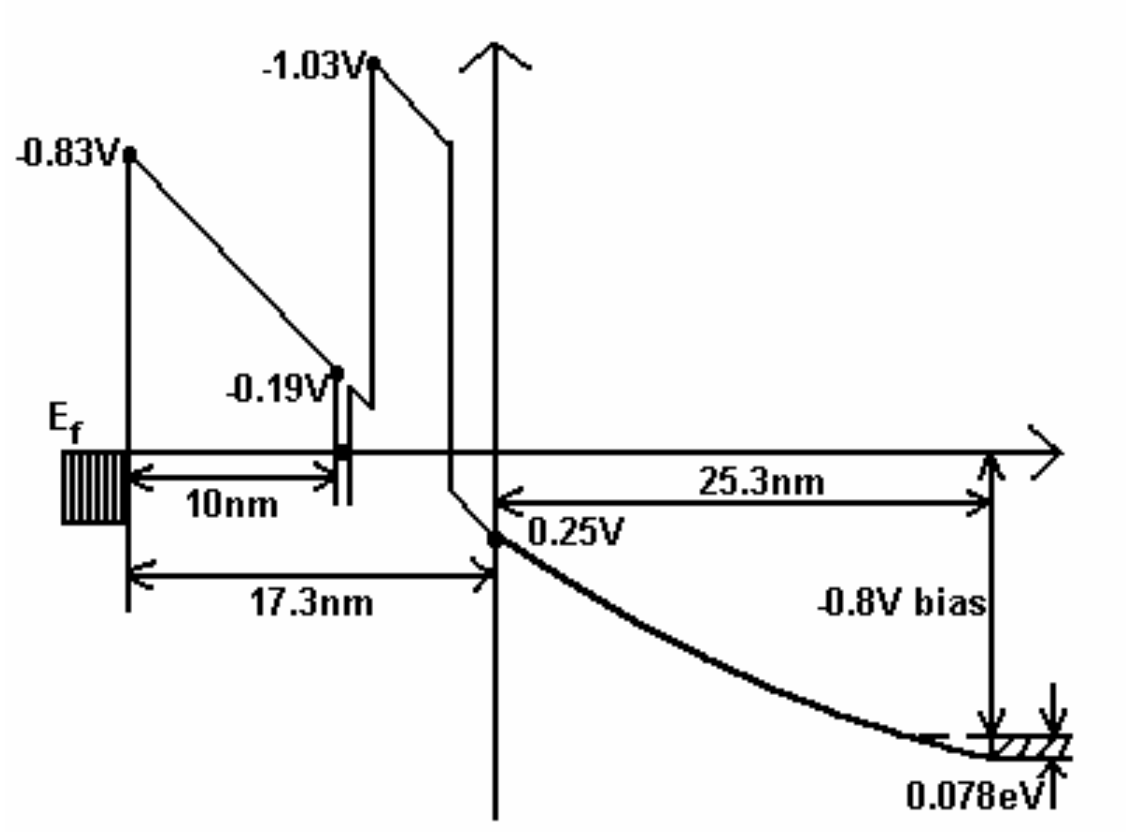

**Figure 6(b)**

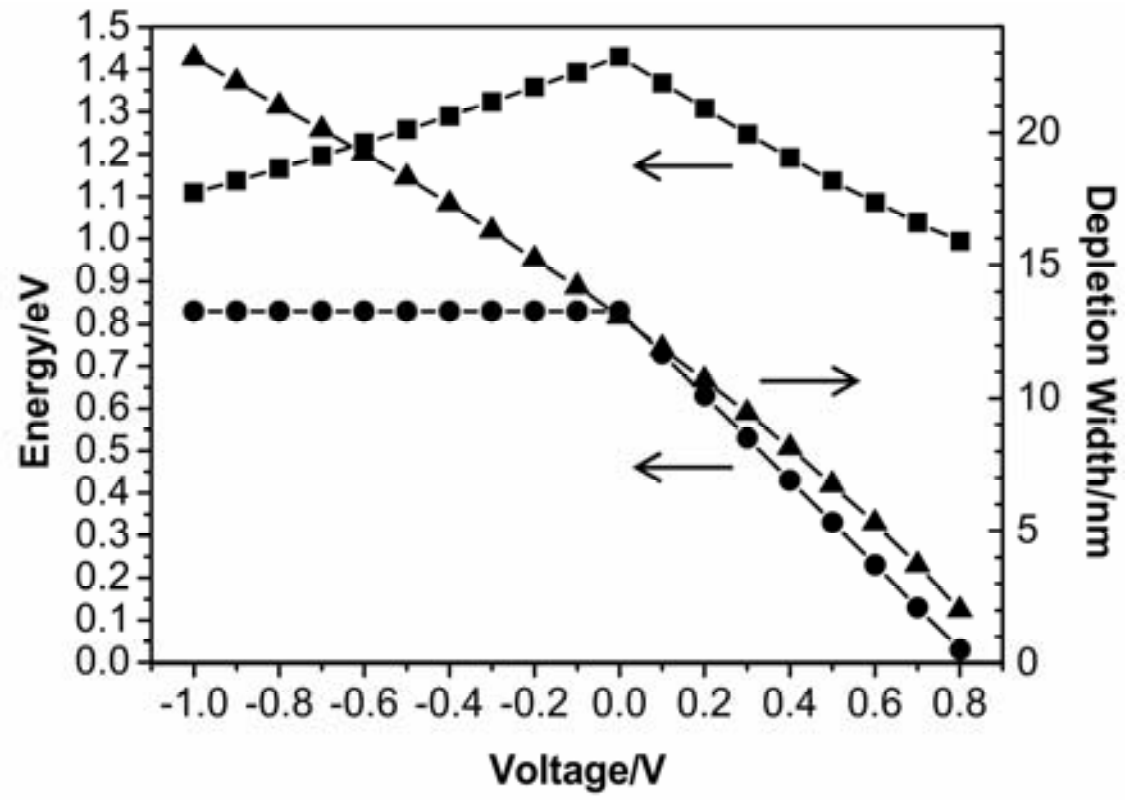
Energy/eV
1.5
1.4
1.3
1.2
1.1
1.0
0.9
0.8
0.7
0.6
0.5
0.4
0.3
0.2
0.1
0.0
-1.0 -0.8 -0.6 -0.4 -0.2 0.0 0.2 0.4 0.6 0.8
Voltage/V
Depletion Width/nm
20
15
10
5
0

**Figure 7**